\documentclass[
*aip,
reprint,
runinaddress,
showpacs,preprintnumbers,
 amsmath,amssymb,
 aps,
pra,
floatfix,
]{revtex4-1}
\usepackage{mathtools}
\usepackage{graphicx}
\usepackage{dcolumn}
\usepackage{bm}
\usepackage{hhline}
\usepackage{color,soul}
\usepackage{float}
\usepackage{mathrsfs}
\usepackage{lipsum}   

\usepackage[normalem]{ulem}
\usepackage[dvipsnames]{xcolor}
\usepackage[sort&compress]{natbib}
\usepackage{xr-hyper}
\usepackage{hyperref}
\usepackage{float}
\hypersetup{
    colorlinks,
    linkcolor={blue},
    citecolor={blue},
    urlcolor={blue}
	}

\newcommand{\mos}{$\rm{MoS_2}$}
\newcommand{\sio}{$\rm{SiO_2/Si}$}

\begin{document}

\title{Contrast inversion in neutral atom microscopy using atomic cluster beams}

\author{Geetika Bhardwaj}%
\affiliation{Tata Institute of Fundamental Research Hyderabad, 36/P Gopanpally, Hyderabad 500046, Telangana, India}

\author{Pranav R. Shirhatti}
\email[Author to whom correspondence should be addressed. \\ e-mail:{\ }]{pranavrs@tifrh.res.in}
\affiliation{Tata Institute of Fundamental Research Hyderabad, 36/P Gopanpally, Hyderabad 500046, Telangana, India}%

\begin{abstract}
\textbf{Abstract:} 
This work explores the possibility of atomic cluster beams as a probe for neutral atom microscopy (NAM) measurements. 
Using a beam of Kr clusters with mean size $\sim$ 10$^4$ atoms/cluster we demonstrate that topographical contrast can be obtained, similar to that in the case of monoatomic beams.
Further, using atomically thin films of \mos{} grown on \sio{} substrate we show that NAM imaging using Kr clusters is also possible in domains where topographical contrast is not expected.
Surprisingly, these images show an inverted contrast pattern when compared to the case of monoatomic beams.
We attempt to understand these observations on the basis of  angular distributions resulting from cluster-surface scattering.   
Finally, we discuss the implications of these results towards achieving a high lateral resolution neutral atom microscope using atomic cluster beams.

\end{abstract}

\maketitle

\section{\label{sec:level1}Introduction}

Using beams of neutral atoms as a probe to image surfaces, also known as Neutral Atom Microscopy (NAM) or Scanning Helium Microscopy (SHeM, in case of He atoms), is an emerging microscopy technique that holds the promise of probing surfaces in a soft manner \cite{allison_2003, witham_NAM_2014}. 
Here, similar to charged particle based methods (like scanning electron microscopy), an atomic beam typically with incident kinetic energy in the range of 10 - 500 meV is made incident on the sample of interest and the scattered atoms are detected in a position-sensitive manner to generate a contrast map (image) of the surface.
One of the major questions in this area of research is to understand contrast generation mechanisms, which is intimately connected to the underlying atom - surface collision dynamics. 
Another major challenge is to achieve high a lateral resolution. 
This is largely constrained by the limited ability to manipulate and control a beam of slow moving neutral atoms.

In the context of microscopy, several schemes to manipulate neutral atom beams have been put forward in the past.
One approach is to focus atomic beams using precisely prepared  surfaces having high reflectivity and appropriate shape, acting as mirrors for atomic beams \cite{holst_mirror1997,holst_ellipsoidalAtomMirror2010, farias_focussing2017, farias_graphenemirror_2011,farias_QuantumStab_AtomMirror2008}. 
Another approach has been to use the wave nature of atoms and focus atomic beams using zone plate structures \cite{toennies_zoneplate1999,allison_2003,koch_imaging_2008, holst_ZonePlate2008,holst_theoretical_model_zoneplate2017}. 
Despite several promising developments in these areas, focusing of atomic beams well beyond 1 $\mu$m spot size remains a scientific and technological challenge.

Arguably, the most successful strategy to date is based on the pinhole design, where a series of apertures are used to collimate the incident atomic beam, consequently reaching a high lateral resolution. 
This design is relatively easier to build and has been widely adopted by several groups \cite{witham, dastoor_NAM_design_2014,dastoor_NAM_Design_RSI2015, bhardwaj2022neutral}.
At present, some of the best images in terms of signal-to-noise ratio \cite{dastoor_NAM_Design_RSI2015} and resolution (sub-micron) \cite{witham_NAM_2014} have been achieved using such pinhole designs.
An excellent overview of different NAM designs, following its systematic development and current state of the art is provided in a recent review article by Palau and coworkers \cite{holst_NAM_reviev2021}.

Despite the promising developments in pinhole based NAM, several challenges still remain to be overcome.
Highest achievable lateral resolution is dictated largely by the dimensions of the final collimation aperture (pinhole). 
Using a smaller pinhole to increase the resolution is necessarily accompanied by a loss of signal-to-noise ratio, as the number of incident particles (and the corresponding scattered signal) decreases. 
Therefore NAM imaging becomes difficult in the region of sub-micron resolution.
Additionally, with pinhole sizes less than 1 $\mu$m, diffraction of incident atoms from the aperture starts becoming significant. 
This causes a lateral spread of the beam, thereby limiting the obtainable resolution for a given working distance. 
A detailed analysis of this situation using numerical simulations has been carried out by Palau and coworkers \cite{holst_theoretical_pinhole}. 
By optimizing the positions of collimating apertures, working distance and accounting for lateral spread of the incident atomic beam caused by diffraction, they estimate that a resolution of 40 nm is achievable under realistic measurement conditions.
It should be noted that such performance remains to be experimentally demonstrated yet.

In this regard, using beams of atomic clusters as a probe offers some interesting possibilities.
Firstly, large atomic density of clusters can compensate for the loss of incident intensity, enabling the use of much smaller pinhole sizes.
At the same time, heavier mass of individual clusters means negligible diffraction effects, even in case of small aperture sizes ($\ll$ 1 $\mu$m). 
It is worth pointing out that a necessary prerequisite to evaluate the potential of these possibilities is to  understand whether the scattering of atomic clusters from surfaces can give rise to contrast maps or not. 
If in case contrast is observed, what is its nature and how does it compare with the usual scenario of monoatomic beams? 
These questions form the subject of our present study.

\begin{figure*}
\centering
\includegraphics[width = 1\linewidth, draft=false]{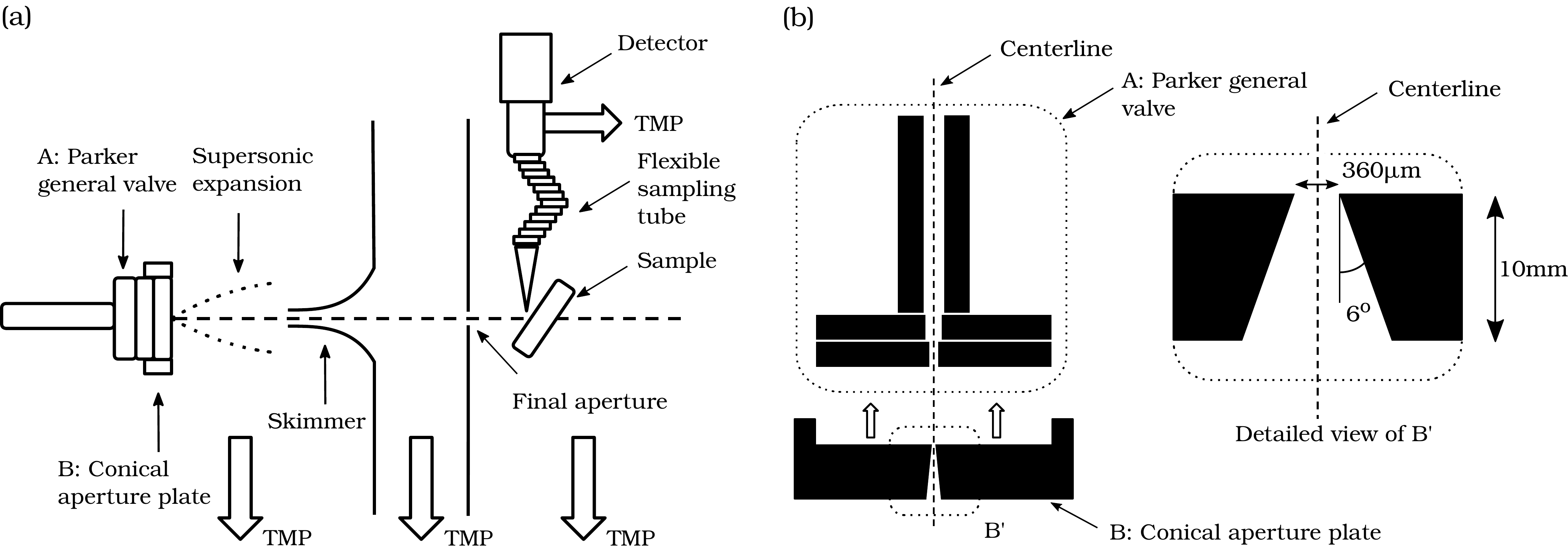}
\caption{ 
(a) Schematic diagram of the overall experimental setup used for NAM measurements. TMP corresponds to turbomolecular pump. 
Supersonic expansion from the nozzle is followed by a 200 $\mu$m diameter opening skimmer and a collimation aperture with diameter 50 $\mu$m.
The collimated beam is made incident from the target sample, placed on a pair of piezo stages (XY), in the scattering chamber.
A fraction of the scattered flux around the specular direction (acceptance angle of $\sim$ 32$^\circ$) enters via the sampling aperture (1 mm opening diameter) into the sampling tube, connected to a quadrupole mass spectrometer with differential pumping arrangement.  
(b) Schematic diagram of pulsed nozzle source and a custom-built conical aperture plate (to aid in cluster formation).
A detailed view of conical aperture plate is shown on the right.}
\label{fig:nam_setup}
\end{figure*}

In the forthcoming sections, we describe the experimental methods used to produce and characterize atomic cluster beams.
Following this we discuss the results of NAM measurements in topographical and beyond topographical contrast regime, where a contrast inversion is seen.
We discuss the possible origins of this unusual contrast inversion and implications of these findings towards developing a high-resolution NAM.

\section{\label{sec:level1}Experimental setup}
\begin{figure}
\centering
\includegraphics[width = 1\linewidth, draft=false]{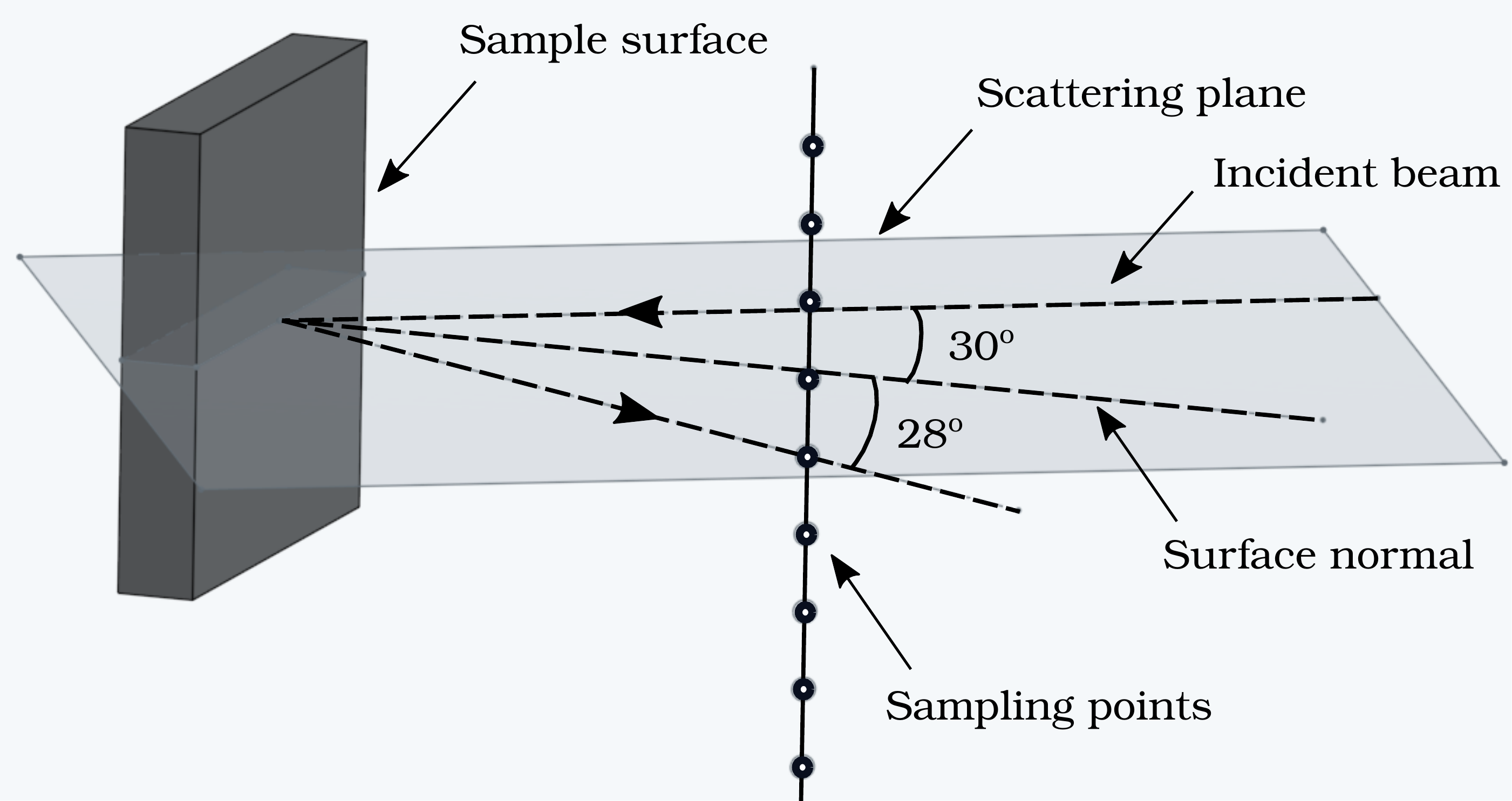}
\caption{Schematic representation of the experimental arrangement used for the angular distribution measurements. Sampling points represent the positions where the measurements were made (perpendicular to the scattering plane) using a flexible sampling tube mounted on a single axis manipulator.
At present, this arrangement limits us to measure only out-of-plane angular distributions.}
\label{fig:ang_dist_geom}
\end{figure}

Experimental setup and sample preparation procedure used in the present work is largely the same as in our previous work \cite{bhardwaj2022neutral}. 
Only specific features essential to understand the present work are described in detail below.
Fig. \ref{fig:nam_setup}\hyperref[fig:nam_setup]{(a)} shows a schematic diagram of the experimental setup used in this work. 
A pulsed atomic beam source comprising of the following components was used: (A) pulsed solenoid valve (Parker 009-1643-900, orifice diameter = 0.5 mm) and (B) a custom-built aperture plate with a 10 mm long conical opening with the diameter of the smaller orifice being 360 $\mu$m and a half opening angle of 6$^\circ$ (Fig. \ref{fig:nam_setup}\hyperref[fig:nam_setup]{(b)}).
This aperture plate was mounted on the front plate of the pulsed valve while keeping their centers aligned.
He or Kr gas, was allowed to expand supersonically from this atomic beam source into the source chamber.   
A 200 $\mu$m skimmer was used to extract the center-line intensity from the gas expansion forming a beam in the first differential chamber. 
Finally a 50 $\mu$m aperture, placed inline, was used to obtain a collimated beam.
The width (full width half maximum, FWHM) of He and Kr beams, measured at a distance of 15 mm (sample plane) from the final collimation aperture using a knife edge scanning method, were observed to be 60 $\mu$m and 56 $\mu$m, respectively. 
These correspond to an angular divergence of $<$ 1 mrad (see SI-1).
For He and Kr beams, based on the pressure changes observed in the detection chamber with the molecular beam on and off (without the sample), we estimate that each gas pulse consists of approximately 10$^{10}$ atoms being incident on the sample.
This corresponds to a flux of $\sim$ 10$^{16}$ atoms/(sec str).
The estimated incidence energies of the pure He and 50\% Kr in He beams are 65 meV and 124 meV, respectively. 
The collimated beam scatters from the sample placed on a movable platform (XY) comprised of two piezoelectric stages  stacked over each other, housed in the detection chamber. 

A 180 mm long flexible stainless steel bellow (inner diameter approximately 3.6 mm) was used as a sampling tube with an orifice of diameter 1 mm drilled at its end.
One end of the sampling tube was mounted on a single axis manipulator, enabling measurement of the scattered signal at different angular positions. 
At present, in this set up the angular distribution measurements, were limited to a plane perpendicular to the scattering plane (Fig. \ref{fig:ang_dist_geom}).
A significant distortion was observed in our attempts (results not shown here) in measuring in-plane angular distributions.
Hence, we restrict ourselves to report and discuss only the out-of-plane scattered distributions in this work.

For NAM measurements, the pulsed valve was driven by a pulse valve driver (IOTA ONE, 060-0001-900, Parker). 
Opening time was set to 25 msec and the repetition rate at 2 Hz. 
Under these conditions, the steady state pressures in the source, first differential  and detection chamber were $\sim$ 3$\times$10$^{-4}$ mbar, $\sim$ 5$\times$10$^{-6}$ mbar, and $\sim$ 3$\times$10$^{-7}$ mbar respectively.
Mass spectrometer (SRS RGA 200) sampling rate was set to 40 Hz and the resulting signal corresponded to pulses of 250-300 ms being detected.
Channeltron voltage was set to -1560 V corresponding to a nominal gain of 1.2$\times$10$^4$.
Dwell time at each sample position (pixel) was set to 2.5 - 3.5 seconds depending on the signal-to-noise ratio and the overall duration of each measurement.

\begin{figure*}[t]
\centering
\includegraphics[width = 1\linewidth, draft=false]{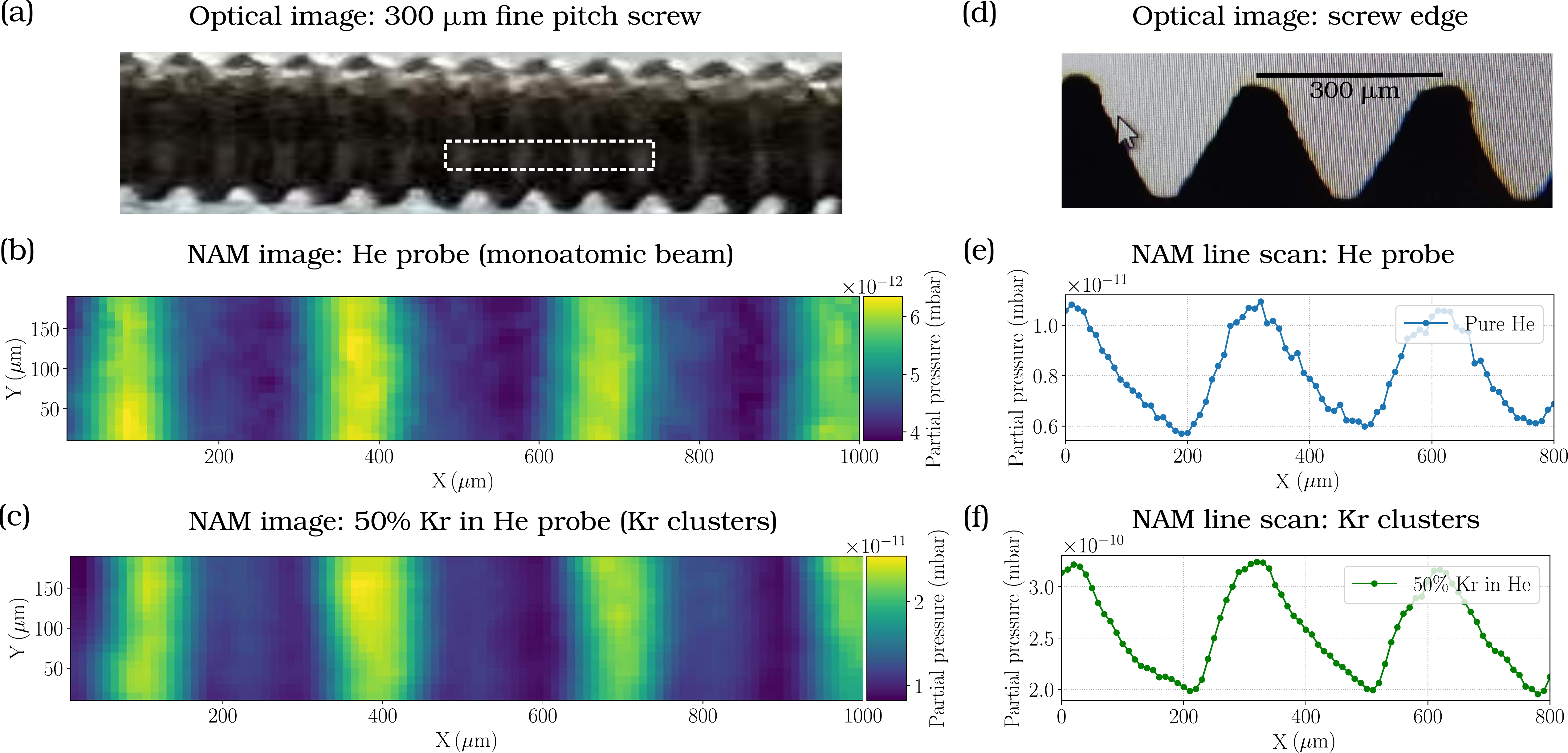}
\caption{
(a) Optical image of a 300 $\mu$m fine-pitch screw (outer diameter of 1.5 mm). A small region marked by a white dashed rectangle was selected to perform the NAM measurements.
(b) and (c) NAM image using monoatomic beam of He atoms and Kr clusters (step-size of 20$\mu$m), respectively. 
(d) Optical image of the edge of screw. 
(e) and (f) Line scan performed along the screw length using a beam of monoatomic He and Kr clusters, respectively (step size 10$\mu$m). 
All the NAM measurements were performed using the custom-built nozzle shown in Fig.\ref{fig:nam_setup}\hyperref[fig:nam_setup]{(b)} with pulse duration set to 25 msec, at a repetition rate of 2Hz.}
\label{fig:nam_screw_panel}
\end{figure*}

The additional aperture plate with a long conical orifice was used to aid in the formation of large clusters.
The relation among average cluster size produced by a nozzle is described empirically by determining the scaling parameter ($\Gamma^*$) as given below \cite{hagena1972cluster,hagena_clusters_ZphysD1987,hagena_ClusterIonSoures_RSI1992,ditmire_laserCluster_1996,ditmire_laserCluster_1998}:

\begin{equation*}
    \Gamma^* = k\frac{(d/\tan\alpha)^{0.85}P_0}{T_0^{2.29}}
\end{equation*}
Here, k is gas dependent condensation parameter \cite{wormer_k_parameter_1989,ditmire_laserCluster_1996,ditmire_laserCluster_1998}, 4 for He, 1700 for Ar, 2900 for Kr, $\alpha$ is the half-angle of the conical aperture (in degrees), d is the orifice diameter (in $\mu$m), $P_0$ is the backing pressure (in mbar), and $T_0$ is the initial gas temperature (in K).

In the present experiments a mixture of 50\% Kr in He was used for NAM measurements with clusters as it resulted in the highest signal observed (see SI-2).
In this case $\Gamma^*$ $\approx$ 1.6$\times 10^4$ (at 6 bar backing pressure).
Under these conditions, based on the previously reported scaling relations among $\Gamma^*$, average cluster size and condensation fraction \cite{hagena_clusters_ZphysD1987, wormer_k_parameter_1989, dorchies_ClusterSizeScaling_2003}, nearly all Kr atoms are expected to be in cluster form with a mean size of 10$^4$ atoms/cluster.
On the other hand, for pure He beams under similar conditions, $\Gamma^*$ is 17 at 2 bar and 86 at 10 bar.  
Therefore, He beams are expected to be largely monoatomic in nature with negligible cluster formation.
Additional measurements to characterize cluster formation were carried out by means of measuring the change in center-line intensity as a function of backing pressure  (see SI-2) and X-ray generation by intense femtosecond laser ionization (see SI-3 and SI-4). 

\begin{figure*}[t]
\centering
\includegraphics[width = 1\linewidth, draft=false]{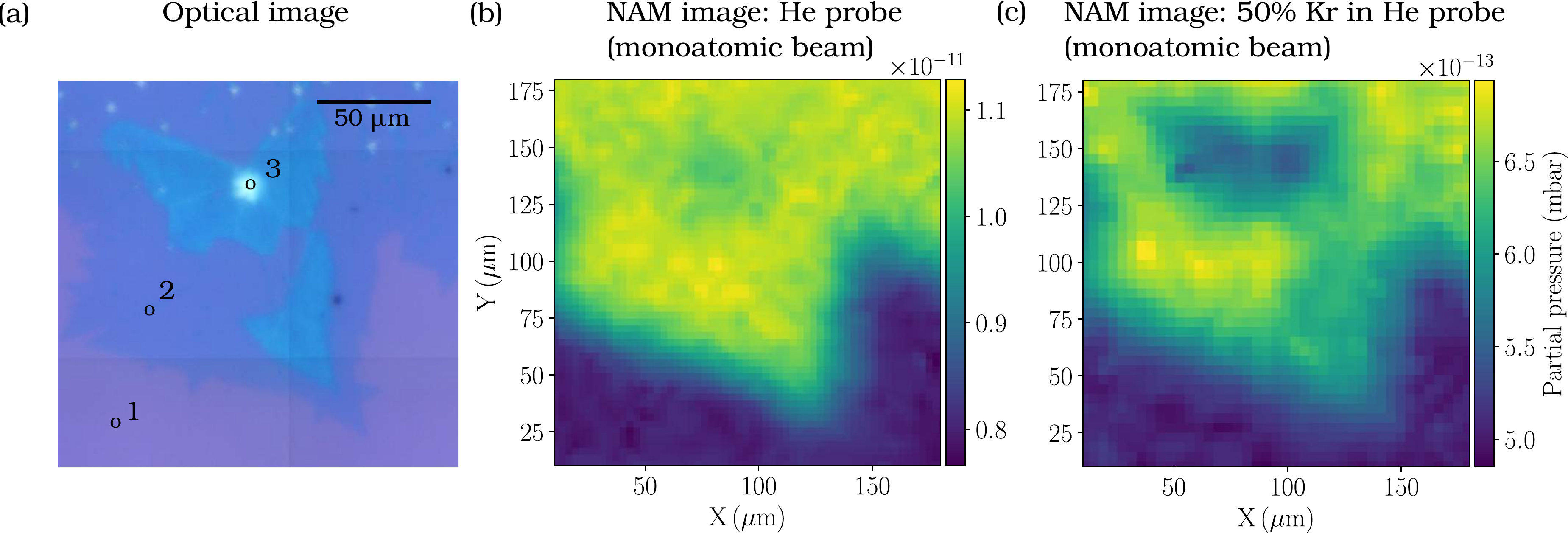}
\caption{
	(a) Optical image of a small portion of MoS$_2$ sample where purple, blue and light-blue colored regions, marked by 1, 2 and 3, correspond to SiO$_2$/Si substrate, thin and thick layers of MoS$_2$ \cite{frindt1966single,bhardwaj2022neutral}. 
	(b) and (c) NAM images obtained using a continuous nozzle with 20 $\mu$m orifice diameter with a monoatomic beam of He and Kr atoms as probe, respectively.
	The scattered flux from MoS$_2$ is consistently higher than SiO$_2$/Si substrate in both cases.
}
\label{fig:mono_He_Kr_panel}
\end{figure*}

Samples of \mos{} grown on \sio{} substrate used in the present study were prepared using chemical vapor deposition method and characterized using optical microscopy and Raman spectroscopy as described previously \cite{bhardwaj2022neutral}.
Features on a typical sample consist of bare substrate, thin (one to three layers) and thick (more than 6 layers). Here one monolayer corresponds to a thickness of 0.65 nm \cite{changgu_MoS2Thickness_raman_2010}. 
These features correspond to the purple, blue and light-blue colors in the optical microscopy images obtained using white light illumination, respectively (see Figs. \ref{fig:mono_He_Kr_panel}  and \ref{fig:cluster_image_panel}).

\section{Results and discussions}

\subsection{\label{sec:level2} Generation of topographical contrast with Kr clusters}

Topographic contrast is the most commonly observed contrast in NAM images \cite{koch_imaging_2008}.
It arises in the regime where incident atoms undergo diffuse scattering upon impact with the surface and spatial features of interest are much larger than the beam spot size.
Under these conditions NAM images closely resemble the geometric features of the sample. 
As far as clusters are concerned, a priori it is not obvious whether simple topographical contrast, commonly seen with monoatomic beams, can be observed or not.
This ambiguity stems from the fact that extent of thermalization of Kr atoms, in the form of large clusters, upon impact with the surface is unknown.

In order to understand if topographic contrast is generated or not, we first image a microscopically rough object (a fine-pitched stainless steel screw, pitch = 300 $\mu$m) using both monoatomic Helium and Krypton cluster beam. 
Features being imaged in this case are much larger than the incident beam size and we can expect to see images largely governed by topographical contrast. 
Figure \ref{fig:nam_screw_panel}\hyperref[fig:nam_screw_panel]{(a)} shows an optical image of the screw.
Panels (b) and (c) show NAM images of a small portion of the screw (marked by a white rectangle in panel (a)), obtained using a beam of Helium and Krypton clusters respectively.
Panel (d) shows an optical image depicting the side view of the screw edge. 
Figures (e) and (f) depict the line profiles  corresponding to the optical image shown in (d), measured using He atoms and Kr clusters, respectively.
Quite clearly, the NAM images obtained using a beam of He (monoatomic) and Kr (clusters) show a one-to-one correspondence with the optical images.
We conclude that large atomic cluster beams are well capable of generating topographical contrast.
We also infer that Kr atoms get thermalized on the surface to a large extent.

These results are of potential interest towards developing a high resolution NAM.
In measurements with monoatomic beams such as He, diffraction of incident atoms from the final aperture sets the ultimate limit to the highest achievable lateral resolution. 
For He atoms, using numerical modelling , this limit has been estimated to be approximately 40 nm \cite{holst_theoretical_pinhole}.
Large atomic clusters such as those used in our experiments, owing to their much higher mass ($\gg$ 10$^4$ times compared to Helium atoms), are expected to behave like classical particles. 
Consequently, the diffraction effects will be negligible even for very small pinhole sizes, providing a route to achieve a higher lateral resolutions compared to monoatomic beams.
In addition, the higher density offered by atomic clusters can lead to much higher incident and scattered signals even with smaller aperture sizes.
We believe that the ultimate limit in this case is likely to be set by the interaction of incident atomic clusters with the edge of small pinhole they are sent through.
At distances where inter-atomic binding energy of the cluster becomes comparable to the interaction energy with atoms constituting the edge of the pinhole, one can expect the clusters to fragment as they travel across the aperture.
Given that these interactions are of Van der Waals type, such forces will be significant only at nm length scales. 
In principle, this can allow nm-sized pinholes to be used for collimation.

\subsection{\label{sec:level2}Beyond topographical contrast: Inverted contrast with Kr clusters}
\begin{figure*}[t]
\centering
\includegraphics[width = 0.85\linewidth, draft=false]{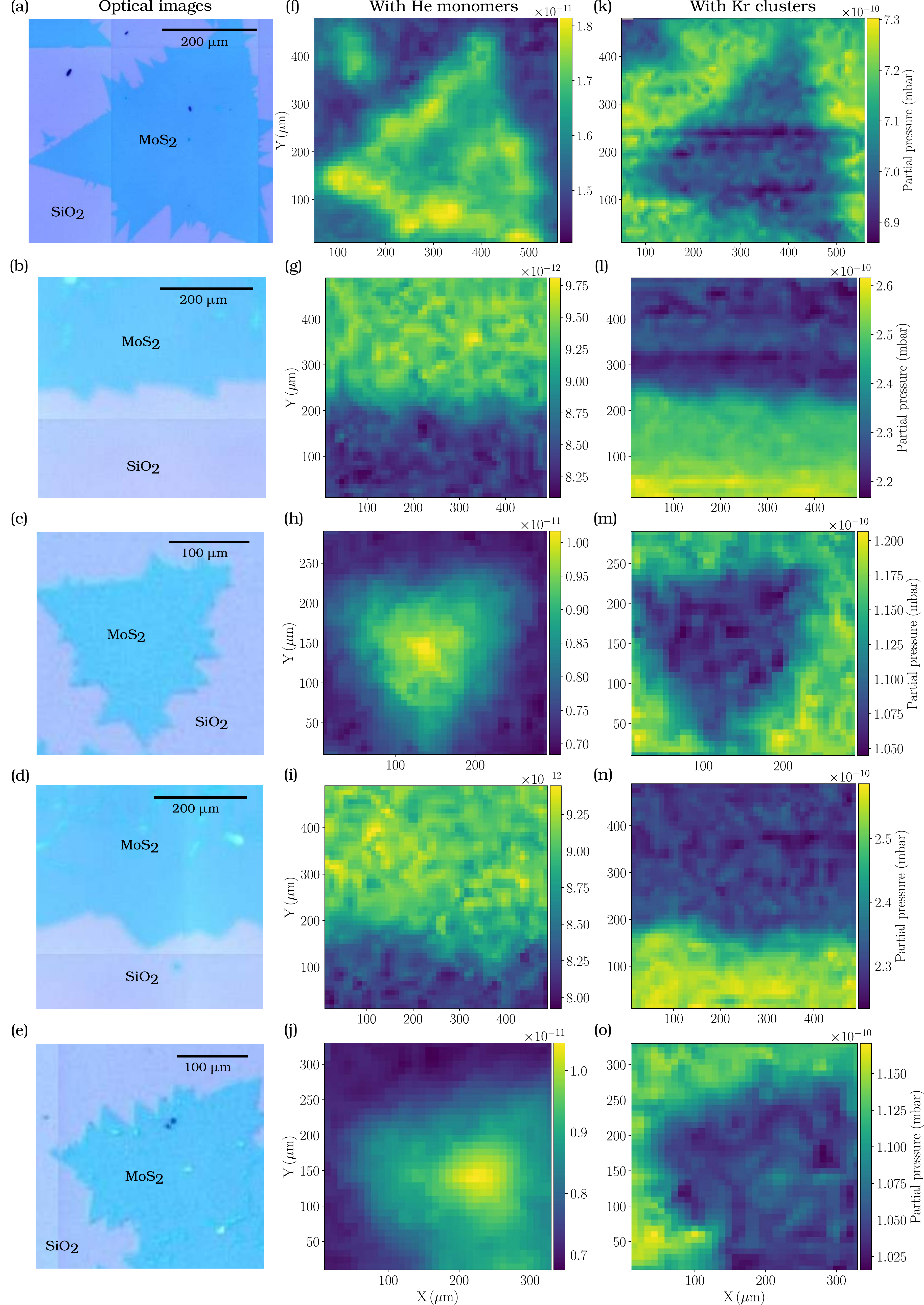}
\caption{
	(a-e) Optical images of five small portions of independently prepared \mos{} films on \sio{} samples. 
	In all the samples, \mos{} and SiO$_2$ regions have been marked. Blue-colored regions contain thin layers $\sim$(0.6-2 nm \cite{bhardwaj2022neutral}) of MoS$_2$. 
	(f-j) NAM images of the same samples using beams of monoatomic He atoms. 
	(k-o) NAM images of the same samples using beams of Kr clusters (mean size $\approx$ 10$^4$ atoms, see text).
	NAM images with both He and Kr clusters show one-to-one correspondence with the optical images.
	Interestingly, images obtained with Kr clusters show an inverted contrast to that compared with He.
	For all NAM measurements shown here, the pulse  duration was set to 25 msec and repetition rate to 2 Hz.}
\label{fig:cluster_image_panel}
\end{figure*}

Regimes beyond topographic contrast correspond to situations where specific details of atom-surface collision play an important role in contrast generation. 
This is unlike the case of topographic contrast resulting from diffuse scattering, where there is no correlation among the incident and final momentum of scattered particles.
As an example, chemical contrast \cite{barr_NatComm2016} has been hypothesized to arise from surface specific inelastic scattering with phonons, leading to a contrast dependent on the surface chemical composition.
Contrast arising due to diffraction of the incident atomic beam as a result of scattering from surfaces with local crystalline order has also been reported recently \cite{jardine_diffraction}.
In our previous work, it was shown that using atom scattering based microscopy, thin films up to a single monolayer of \mos{} on \sio{} can be successfully imaged using a 20-30 $\mu$m sized beam of He and/or Kr atoms as an incident probe.
Further it was also observed that contrast was degrading with incident energy \cite{bhardwaj2022neutral}. 
These results point towards the fact that contrast mechanisms beyond simple topographical in nature are at play.
Here, we investigate whether NAM imaging of atomically thin layers of MoS$_2$, in the beyond topographic contrast regime, is possible with a beam of Kr clusters or not.

\begin{figure*}[t]
\centering
\includegraphics[width = 1\linewidth, draft=false]{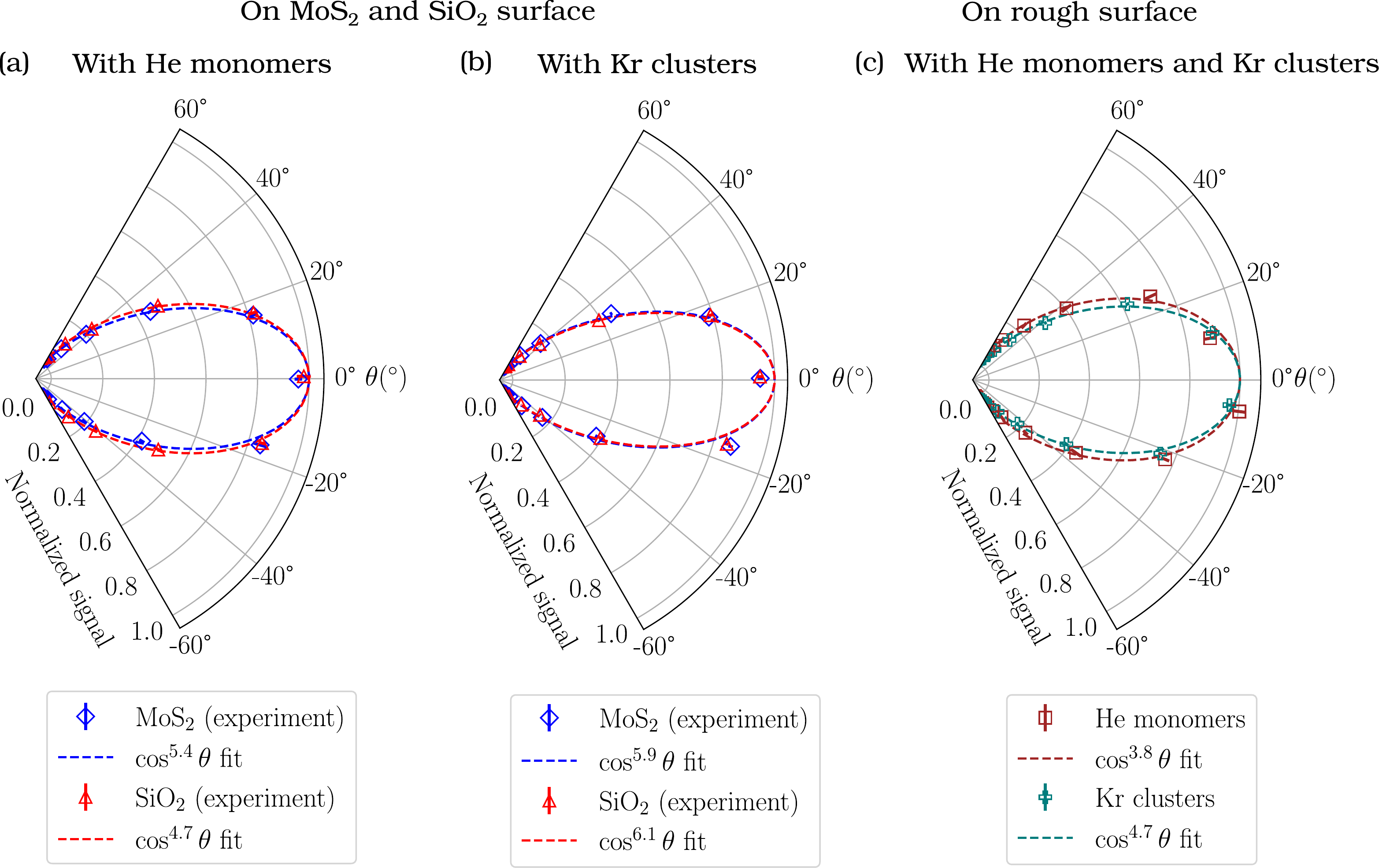}
\caption{
	(a) and (b) Angle-resolved scattered flux from \mos{} and \sio{} substrate using a beam of monoatomic He atoms and Kr clusters. 
	For He, the distribution obtained from MoS$_2$ surface is slightly narrower than that of SiO$_2$ surface, consistent with the observed contrast pattern.
	In case of Kr clusters, both the distributions (b) look very similar.
	Angular distribution obtained from a rough stainless steel surface is shown in (c) for comparison.
	The angular position of 0$^\circ$ corresponds to the surface normal. 
}
\label{fig:ang_dist}
\end{figure*}

Measurements with monoatomic beam of He and Kr atoms, produced with a 20 $\mu$m continuous nozzle, can be seen in Fig. \ref{fig:mono_He_Kr_panel}. 
Panel (a) shows an optical image of a small portion of $\rm MoS_2$ grown on $\rm SiO_2$/Si substrate and the corresponding NAM images obtained with He and Kr beams are depicted in (b) and (c), respectively. 
The scattered flux of He and Kr from \mos{} is consistently higher as compared to \sio{} in both cases and a clear one-to-one correspondence is seen. 
Such kind of contrast generation with the underlying possible reasons have been discussed in our previous work \cite{bhardwaj2022neutral}. 

Measurements performed with a beam of Kr clusters and a monoatomic beam of He are shown in Fig. \ref{fig:cluster_image_panel}. 
Panels \ref{fig:cluster_image_panel}\hyperref[fig:cluster_image_panel]{(a - e)} show optical images of  small portions of independently prepared samples of \mos{} films on \sio{} substrate. 
NAM measurements using He atom beam are shown in Fig. \ref{fig:cluster_image_panel}\hyperref[fig:cluster_image_panel]{(f - j)}. 
Here we observe the expected contrast pattern as seen previously with monoatomic beams (Fig. \ref{fig:mono_He_Kr_panel}), i.e. higher scattered flux from \mos{} surface as compared to \sio{}.
Figs. \ref{fig:cluster_image_panel}\hyperref[fig:cluster_image_panel]{(g - i)} show the images obtained using Kr clusters.
Again, a one-to-one correspondence with the optical images is seen, but interestingly, contrast patterns appear inverted.
Here, a larger signal for scattered Kr is obtained from \sio{} substrate as compared to regions covered with \mos{}.
Quite clearly, scattering of large Kr clusters does not simply lead to diffuse scattering alone and also seems to be sensitive to the surface characteristics.
Observation of such a contrast inversion merits further discussion.

Previous studies on scattering of atomic and molecular clusters from surfaces, using experimental \cite{gspann_ClusterReflectionSS_1974, bernasek_Cluster_Fe_scattering_1988, vach_ArgonCluster_graphite_1992, vach_N2Clusters_Graphite_1999}
and computational methods \cite{tully_ArClusters_Pt_1989, pettersson_ArClusters_Pt_1997},  offer valuable insights.
These studies span a range of atomic and molecular clusters (such as He, H$_2$, N$_2$ and Ar) and different surfaces ranging from microscopically rough to atomically flat single crystals. 
A common feature arising in all these works is that for large clusters, angular distributions of the scattered particles show a prominent peak at angles much greater than specular direction (supraspecular).
Further, it has also been reported that on rough surfaces (prepared by ion sputtering of atomically flat, clean single crystal surfaces) this pronounced supraspecular scattering becomes weaker and the angular distributions tend towards diffuse scattering (peaking at surface normal) \cite{bernasek_Cluster_Fe_scattering_1988}.

This general aspect of cluster scattering from surfaces seems significant as far as our observation of contrast inversion is concerned.
For monoatomic beams, the scattered flux is expected to peak close to the specular direction in case of elastic scattering or towards the surface normal, in case of a large diffuse scattering component.
On the other hand, for large Kr clusters (n $\sim$ 10$^4$), the majority of the scattered flux is likely to exit at large angles from surface normal, in supraspecular direction. Given that our sampling aperture is placed near specular direction, a large fraction of the supraspecular scattered atoms will not be captured by the detector.
Our previous study has shown that at an atomic scale, \mos{} surfaces are generally smoother than the \sio{} substrate \cite{bhardwaj2022neutral}.
Given the above points, we hypothesize that on \mos{} surface, supraspecular scattering leads to relatively fewer atoms entering the collection aperture placed along the specular direction.
At the same time, the increased roughness on \sio{} substrate can lead to more diffuse scattering like behavior, resulting in a relatively higher scattered signal being detected.

Besides the above possibility, other systematic differences could also be at play here.
\mos{} films are bound to the lower layers or \sio{} substrate by relatively weak Van der Waals interaction. 
As a result, \mos{} surface acts as a relatively softer landing site for Kr clusters.
Large Kr clusters having sizable incidence energy compared to its monoatomic counterparts, when scattered from a relatively softer surface (compared to \sio{}) can lead to a broad scattered distribution.
Consequently, a lower flux will be measured by the detector placed in specular direction compared to scattering from a relatively rigid \sio{} surface, consistent with the inverted contrast observed in our measurements.
The effect of rigidity of thin films on the scattered angular distributions have been studied by Taleb and co-workers \cite{farias_interlayerForces_2018} by studying the heavy vs. light atom scattering from graphene surface in the case of weak and strong interaction with the substrate.
They observed that in case of weak interaction with substrate (Gr/Ir(111)), neon atoms scattered largely inelastically leading to a broad angular distribution. 
On the other hand, scattering from a more strongly bound substrate such as Gr/Ni(111), sharp diffraction peaks were obtained.
 
In order to understand these points better, we resort to angle-resolved measurements.
Figs. \ref{fig:ang_dist}\hyperref[fig:ang_dist]{(a)} and \hyperref[fig:ang_dist]{(b)} show a comparison of the angular distributions resulting from scattering on \sio{} and \mos{} surfaces, obtained using He (monoatomic) and Kr (clusters), respectively. 
Angular distributions from a microscopically rough surface (stainless steel screw, shown in Fig. \ref{fig:nam_screw_panel} are also shown in (c) for the sake of comparison.
It should be noted that these distributions were measured for an out-of-plane scattering configuration (see Fig. \ref{fig:ang_dist_geom}) and are not corrected for any distortions resulting from the measurement configuration. 
Nonetheless, a systematic comparison of the relative changes based on these results is still possible.

A common feature observed in the case of \mos{} and \sio{} is that angular distributions obtained using monoatomic He are rather broad and only slightly narrower than that observed from a rough surface. 
This indicates that a substantial fraction of the atoms undergo diffuse scattering. 
This is expected since our samples were placed in a vacuum chamber operating at a base pressure of 3$\times$10$^{-7}$ mbar and not true ultrahigh vacuum conditions. Further, no \textit{in-situ} sample cleaning was done.
Under these conditions, a significant amount of adsorbates will be present on the surface, leading to a large diffuse scattering component.
For He beam, width of the distribution in case of \mos{} surface, is somewhat narrower than \sio{} substrate, consistent with the contrast observed in NAM images.
Angular distributions obtained with Kr clusters appear very similar for \mos{}  and \sio{}.
Given that these distributions almost look same and the fact that we see an inverted contrast pattern, we infer that the changes are largely occurring in the in-plane scattering distributions. This is also expected from previous studies showing supraspecular scattering.
A clear answer to these questions can be obtained by in-plane angle-resolved measurements, which is currently unavailable in our setup.
An upgraded version with provision for measuring both, in-plane and out-of-plane angular distributions with the aid of a combined rotation and linear manipulator along with sample heating capability (for removing weakly bound adsorbates) is being designed in our lab for future studies.

\section{Concluding Remarks}

In this work we have demonstrated that NAM imaging is possible using a beam of Kr atom clusters and the well-known topographic contrast can be obtained for rough surfaces.
Using samples of atomically thin films of \mos{} grown on \sio{} substrate, we have shown that NAM imaging with Kr clusters is possible even in the regime of beyond simple topographical contrast.
Interestingly, here we observe a contrast inversion compared to the similar measurements made with monoatomic beams. 

Importantly, the results presented here clearly establish that NAM imaging can be done with atomic clusters as well.
To the best of our knowledge, this possibility has not been explored previously.
These results point toward two interesting possibilities in the direction of developing a high lateral resolution NAM. 
Firstly, the higher atomic density of clusters can be exploited to obtain high incident beam flux which can in turn allow the use of smaller pinholes leading to higher lateral resolution.
Secondly, in the case of atomic clusters, owing to their much higher mass compared to their monoatomic counterparts, the problem of lateral spread caused by diffraction from small pinholes is expected to be negligible.
A direct consequence of both the above points is that smaller pinhole sizes of the order of cluster sizes (few nm) can be used, providing a possible route towards realizing a high lateral resolution (few nm) neutral atom microscope.

A systematic exploration of these possibilities will be needed especially to understand the maximum centerline intensity obtainable in case of atomic cluster beams \cite{korobeishchikov_MeanCluster_2017}. 
Here, the role of mass focusing effect \cite{knuth_massFocussing_1976} towards enhancing the density of clusters along the centerline and the counter-acting effect of warming up of the beam to reduce the centerline intensity need to be understood well.
Further, these results also suggest an interesting possibility of developing a guided negatively charged cluster ion source with subsequent neutralization. 
Using this approach, a tightly focused beam of low-energy neutral atomic clusters can be generated.
Such focused neutral atomic cluster beam sources can be of interest for developing a high lateral resolution NAM.

\section*{Supplementary Information}
\begin{itemize}
	\item SI-1: Incident beam width estimation
	\item SI-2: Variation in scattered signal with backing pressure
	\item SI-3: Experimental setup to verify cluster formation using X-ray generation
	\item SI-4: X-ray spectra
\end{itemize}

\section*{Data Availability}
All relevant data related to the current study are available from the corresponding author upon reasonable request.

\section*{Acknowledgements}
This work was partly supported by intramural funds at TIFR Hyderabad from the Department of Atomic Energy and Scientific and Engineering Research Board, Department of Science and Technology (grant numbers: CRG/2020/003877 and ECR/2018/001127). 
We thank T. N. Narayanan and Dipak Maity for sample preparation and characterization. Vandana Sharma (IIT-Hyderabad) for providing vacuum manipulator, M. Krishnamurthy and lab members for providing a turbo molecular pump and help with the femtosecond laser ionization and X-ray emission measurements, Rakesh Moodika for fabricating the sampling aperture and parts for the movable sampling tube assembly.

\section*{Author Contributions} GB designed and characterized the NAM and the cluster beam experimental setup with inputs from PRS. GB performed the measurements and analyzed the data. PRS conceived the idea and provided conceptual inputs. GB and PRS discussed the results and prepared the manuscript.

\bibliographystyle{unsrt}
\bibliography{references.bib}

\begin{thebibliography}{10}

\bibitem{allison_2003}
D.~A. MacLaren, B.~Holst, D.~J. Riley, and W.~Allison.
\newblock Focusing elements and design considerations for a scanning {H}elium
  microscope {(SHeM)}.
\newblock {\em Surface Review and Letters}, 10(02n03):249--255, 2003.

\bibitem{witham_NAM_2014}
P.~Witham and E.~Sánchez.
\newblock Exploring neutral atom microscopy: {Exploring} neutral atom
  microscopy.
\newblock {\em Crystal Research and Technology}, 49(9):690--698, September
  2014.

\bibitem{holst_mirror1997}
B.~Holst and W.~Allison.
\newblock An atom-focusing mirror.
\newblock {\em Nature}, 390(6657):244--244, 1997.

\bibitem{holst_ellipsoidalAtomMirror2010}
K.~Fladischer, H.~Reingruber, T.~Reisinger, V.~Mayrhofer, W.~E. Ernst, A.~E.
  Ross, D.~A. MacLaren, W.~Allison, D.~Litwin, J.~Galas, et~al.
\newblock An ellipsoidal mirror for focusing neutral atomic and molecular
  beams.
\newblock {\em New journal of Physics}, 12(3):033018, 2010.

\bibitem{farias_focussing2017}
Gloria Anemone, Amjad~Al Taleb, Sabrina~D. Eder, Bodil Holst, and Daniel
  Farías.
\newblock Flexible thin metal crystals as focusing mirrors for neutral atomic
  beams.
\newblock {\em Physical Review B}, 95(20):205428, May 2017.

\bibitem{farias_graphenemirror_2011}
P.~Sutter, M.~Minniti, P.~Albrecht, D.~Farías, R.~Miranda, and E.~Sutter.
\newblock A high-reflectivity, ambient-stable graphene mirror for neutral
  atomic and molecular beams.
\newblock {\em Applied Physics Letters}, 99(21):211907, November 2011.

\bibitem{farias_QuantumStab_AtomMirror2008}
D.~Barredo, F.~Calleja, P.~Nieto, J.~J. Hinarejos, G.~Laurent, A.~L.
  {V{\'a}zquez de Parga}, D.~Far{\'\i}as, and R.~Miranda.
\newblock A quantum-stabilized mirror for atoms.
\newblock {\em Advanced materials}, 20(18):3492--3497, 2008.

\bibitem{toennies_zoneplate1999}
R.~B. Doak, R.~E. Grisenti, S.~Rehbein, G.~Schmahl, J.~P. Toennies, and
  C.~W{\"o}ll.
\newblock Towards realization of an atomic de {B}roglie microscope: {H}elium
  atom focusing using fresnel zone plates.
\newblock {\em Physical review letters}, 83(21):4229, 1999.

\bibitem{koch_imaging_2008}
M.~Koch, S.~Rehbein, G.~Schmahl, T.~Reisinger, G.~Bracco, W.~E. Ernst, and
  B.~Holst.
\newblock Imaging with neutral atoms—a new matter-wave microscope.
\newblock {\em Journal of Microscopy}, 229(1):1--5, January 2008.

\bibitem{holst_ZonePlate2008}
T.~Reisinger and B.~Holst.
\newblock Neutral atom and molecule focusing using a fresnel zone plate.
\newblock {\em Journal of Vacuum Science \& Technology B: Microelectronics and
  Nanometer Structures Processing, Measurement, and Phenomena},
  26(6):2374--2379, 2008.

\bibitem{holst_theoretical_model_zoneplate2017}
A.~S. Palau, G.~Bracco, and B.~Holst.
\newblock Theoretical model of the {H}elium zone plate microscope.
\newblock {\em Physical Review A}, 95(1):013611, 2017.

\bibitem{witham}
P.~Witham and E.~S{\'a}nchez.
\newblock A simple approach to neutral atom microscopy.
\newblock {\em Review of Scientific Instruments}, 82(10):103705, 2011.

\bibitem{dastoor_NAM_design_2014}
M.~Barr, A.~Fahy, A.~Jardine, J.~Ellis, D.~Ward, D.A. MacLaren, W.~Allison, and
  P.C. Dastoor.
\newblock A design for a pinhole scanning helium microscope.
\newblock {\em Nuclear Instruments and Methods in Physics Research Section B:
  Beam Interactions with Materials and Atoms}, 340:76--80, December 2014.

\bibitem{dastoor_NAM_Design_RSI2015}
A.~Fahy, M.~Barr, J.~Martens, and P.~C. Dastoor.
\newblock A highly contrasting scanning helium microscope.
\newblock {\em Review of Scientific Instruments}, 86(2):023704, February 2015.

\bibitem{bhardwaj2022neutral}
Geetika Bhardwaj, Krishna~Rani Sahoo, Rahul Sharma, Parswa Nath, and Pranav~R
  Shirhatti.
\newblock Neutral-atom-scattering-based mapping of atomically thin layers.
\newblock {\em Physical Review A}, 105(2):022828, 2022.

\bibitem{holst_NAM_reviev2021}
Adrià~Salvador Palau, Sabrina~Daniela Eder, Gianangelo Bracco, and Bodil
  Holst.
\newblock Neutral {Helium} {Microscopy} ({SHeM}): {A} {Review}.
\newblock {\em arXiv:2111.12582 [physics]}, December 2021.
\newblock arXiv: 2111.12582.

\bibitem{holst_theoretical_pinhole}
A.~S. Palau, G.~Bracco, and B.~Holst.
\newblock Theoretical model of the {H}elium pinhole microscope.
\newblock {\em Physical Review A}, 94(6):063624, 2016.

\bibitem{hagena1972cluster}
O.~F. Hagena and W.~Obert.
\newblock Cluster formation in expanding supersonic jets: Effect of pressure,
  temperature, nozzle size, and test gas.
\newblock {\em The Journal of Chemical Physics}, 56(5):1793--1802, 1972.

\bibitem{hagena_clusters_ZphysD1987}
O.~F. Hagena.
\newblock Condensation in free jets: {Comparison} of rare gases and metals.
\newblock {\em Zeitschrift f\"ur Physik D Atoms, Molecules and Clusters},
  4(3):291--299, September 1987.

\bibitem{hagena_ClusterIonSoures_RSI1992}
O.~F. Hagena.
\newblock Cluster ion sources.
\newblock {\em Review of scientific instruments}, 63(4):2374--2379, 1992.

\bibitem{ditmire_laserCluster_1996}
T.~Ditmire, T.~Donnelly, A.~M. Rubenchik, R.~W. Falcone, and M.~D. Perry.
\newblock Interaction of intense laser pulses with atomic clusters.
\newblock {\em Physical Review A}, 53(5):3379--3402, May 1996.

\bibitem{ditmire_laserCluster_1998}
T.~Ditmire, E.~Springate, J.~W.~G. Tisch, Y.~L. Shao, M.~B. Mason, N.~Hay,
  J.~P. Marangos, and M.~H.~R. Hutchinson.
\newblock Explosion of atomic clusters heated by high-intensity femtosecond
  laser pulses.
\newblock page~14, 1998.

\bibitem{wormer_k_parameter_1989}
J.~Wörmer, V.~Guzielski, J.~Stapelfeldt, and T.~Möller.
\newblock Fluorescence excitation spectroscopy of xenon clusters in the {VUV}.
\newblock {\em Chemical Physics Letters}, 159(4):321--326, July 1989.

\bibitem{dorchies_ClusterSizeScaling_2003}
F.~Dorchies, F.~Blasco, T.~Caillaud, J.~Stevefelt, C.~Stenz, A.~S. Boldarev,
  and V.~A. Gasilov.
\newblock Spatial distribution of cluster size and density in supersonic jets
  as targets for intense laser pulses.
\newblock {\em Physical Review A}, 68(2):023201, August 2003.

\bibitem{changgu_MoS2Thickness_raman_2010}
C.~Lee, H.~Yan, L.~E. Brus, T.~F Heinz, J.~Hone, and S.~Ryu.
\newblock Anomalous lattice vibrations of single-and few-layer {MoS$_2$}.
\newblock {\em ACS nano}, 4(5):2695--2700, 2010.

\bibitem{frindt1966single}
R.~F. Frindt.
\newblock Single crystals of mos2 several molecular layers thick.
\newblock {\em Journal of Applied Physics}, 37(4):1928--1929, 1966.

\bibitem{barr_NatComm2016}
M.~Barr, A.~Fahy, J.~Martens, A.~P. Jardine, D.~J. Ward, J.~Ellis, W.~Allison,
  and P.~C. Dastoor.
\newblock Unlocking new contrast in a scanning {H}elium microscope.
\newblock {\em Nature communications}, 7(1):1--5, 2016.

\bibitem{jardine_diffraction}
M.~Bergin, S.~M. Lambrick, H.~Sleath, D.~J. Ward, J.~Ellis, and A.~P. Jardine.
\newblock Observation of diffraction contrast in scanning {H}elium microscopy.
\newblock {\em Scientific reports}, 10(1):1--8, 2020.

\bibitem{gspann_ClusterReflectionSS_1974}
J.~Gspann and G.~Krieg.
\newblock Reflection of clusters of helium, hydrogen, and nitrogen as function
  of the reflector temperature.
\newblock {\em The Journal of Chemical Physics}, 61(10):4037--4047, November
  1974.

\bibitem{bernasek_Cluster_Fe_scattering_1988}
R.~J. Holland, G.~Q. Xu, J.~Levkoff, A.~Robertson, and S.~L. Bernasek.
\newblock Experimental studies of the dynamics of nitrogen van der {Waals}
  cluster scattering from metal surfaces.
\newblock {\em The Journal of Chemical Physics}, 88(12):7952--7963, June 1988.

\bibitem{vach_ArgonCluster_graphite_1992}
M.~Châtelet, A.~De~Martino, J.~Pettersson, F.~Pradère, and H.~Vach.
\newblock Argon cluster scattering from a graphite surface.
\newblock {\em Chemical Physics Letters}, 196(6):563--568, August 1992.

\bibitem{vach_N2Clusters_Graphite_1999}
A.~De~Martino, M.~Châtelet, F.~Pradère, E.~Fort, and H.~Vach.
\newblock Experimental investigation of large nitrogen cluster scattering from
  graphite: {Translational} and rotational distributions of evaporated {N$_2$}
  molecules.
\newblock {\em The Journal of Chemical Physics}, 111(15):7038--7046, October
  1999.

\bibitem{tully_ArClusters_Pt_1989}
Guo-Qin Xu, R.~J. Holland, and Steven~L. Bernasek.
\newblock Dynamics of cluster scattering from surfaces.
\newblock {\em J. Chem. Phys.}, 90:8, 1989.

\bibitem{pettersson_ArClusters_Pt_1997}
Marcus Svanberg, Nikola Marković, and Jan~B.C. Pettersson.
\newblock Scattering of large argon clusters from a {Pt}(111) surface with low
  collision velocities.
\newblock {\em Chemical Physics}, 220(1-2):137--153, July 1997.

\bibitem{farias_interlayerForces_2018}
Amjad Al~Taleb, Gloria Anemone, Rodolfo Miranda, and Daniel Farías.
\newblock Characterization of interlayer forces in {2D} heterostructures using
  neutral atom scattering.
\newblock {\em 2D Materials}, 5(4):045002, July 2018.

\bibitem{korobeishchikov_MeanCluster_2017}
N.~G. Korobeishchikov, M.~A. Roenko, and G.~I. Tarantsev.
\newblock Mean {Gas} {Cluster} {Size} {Determination} from {Cluster} {Beam}
  {Cross}-{Section}.
\newblock {\em Journal of Cluster Science}, 28(5):2529--2547, September 2017.

\bibitem{knuth_massFocussing_1976}
P.~K. Sharma, E.~L. Knuth, and W.~S. Young.
\newblock Species enrichment due to {Mach}‐number focusing in a
  molecular‐beam mass‐spectrometer sampling system.
\newblock {\em The Journal of Chemical Physics}, 64(11):4345--4351, June 1976.

\end{thebibliography}

\end{document}


\title{Contrast inversion in neutral atom microscopy using atomic cluster beams - Supplementary Information}

\author{Geetika Bhardwaj}%
\affiliation{Tata Institute of Fundamental Research Hyderabad, 36/P Gopanpally, Hyderabad 500046, Telangana, India}

\author{Pranav R. Shirhatti}
\email[Author to whom correspondence should be addressed. \\ e-mail:{\ }]{pranavrs@tifrh.res.in}
\affiliation{Tata Institute of Fundamental Research Hyderabad, 36/P Gopanpally, Hyderabad 500046, Telangana, India}%

\maketitle

\section*{SI-1: Incident beam width estimation}
\begin{figure}[H]
\centering
\includegraphics[width = 0.85\linewidth, draft=false]{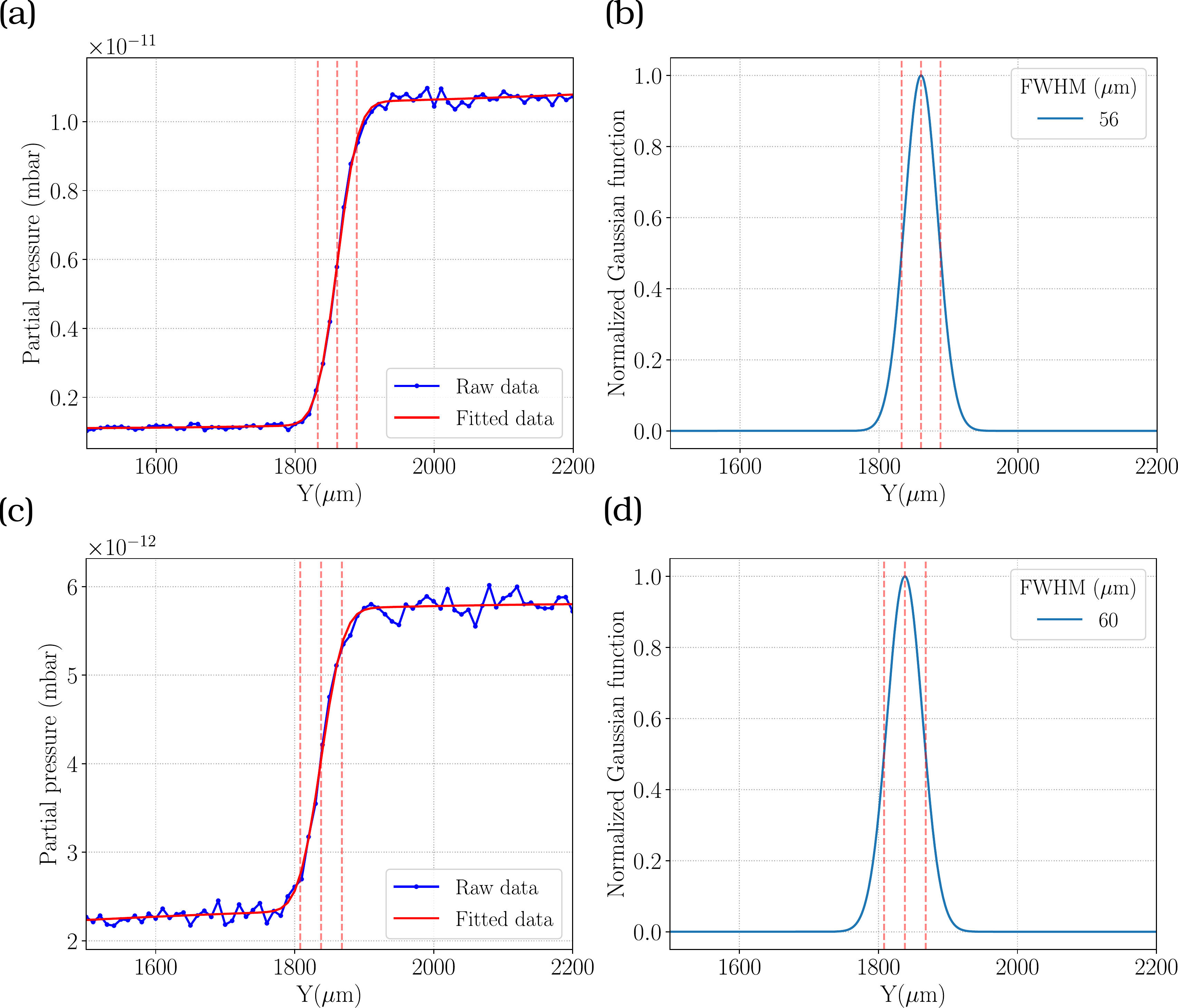}
\caption{
	Incident beam-width estimation on the sample plane using knife-edge scan method. 
	(a) Beam width measurement for Kr in 50\% Kr + 50\% He mixture. Blue curve shows the signal of Kr observed as a function of the position of knife-edge (razor blade) with step size of 10$\mu$m. Red curve shows the best fit, using a model based on a step function convoluted with a Gaussian. 
	(b) Modeled beam profile with the parameters obtained by fitting with a FWHM of 56 $\mu$m. 
	(c) and (d) depict the same for pure He (monoatomic) resulting in a beam width estimate of 60 $\mu$m (FWHM).
}
\label{fig:beam_width_panel}
\end{figure}

Beam-width estimation of monoatomic He beam and a beam of Kr clusters (using 50\% mixture) were measured using knife-edge scan method. These results are shown in Fig. \ref{fig:beam_width_panel}.

\newpage 
\section*{SI-2: Variation in Scattered signal with backing pressure}

\begin{figure}[H]
\centering
\includegraphics[width = 0.9\linewidth, draft=false]{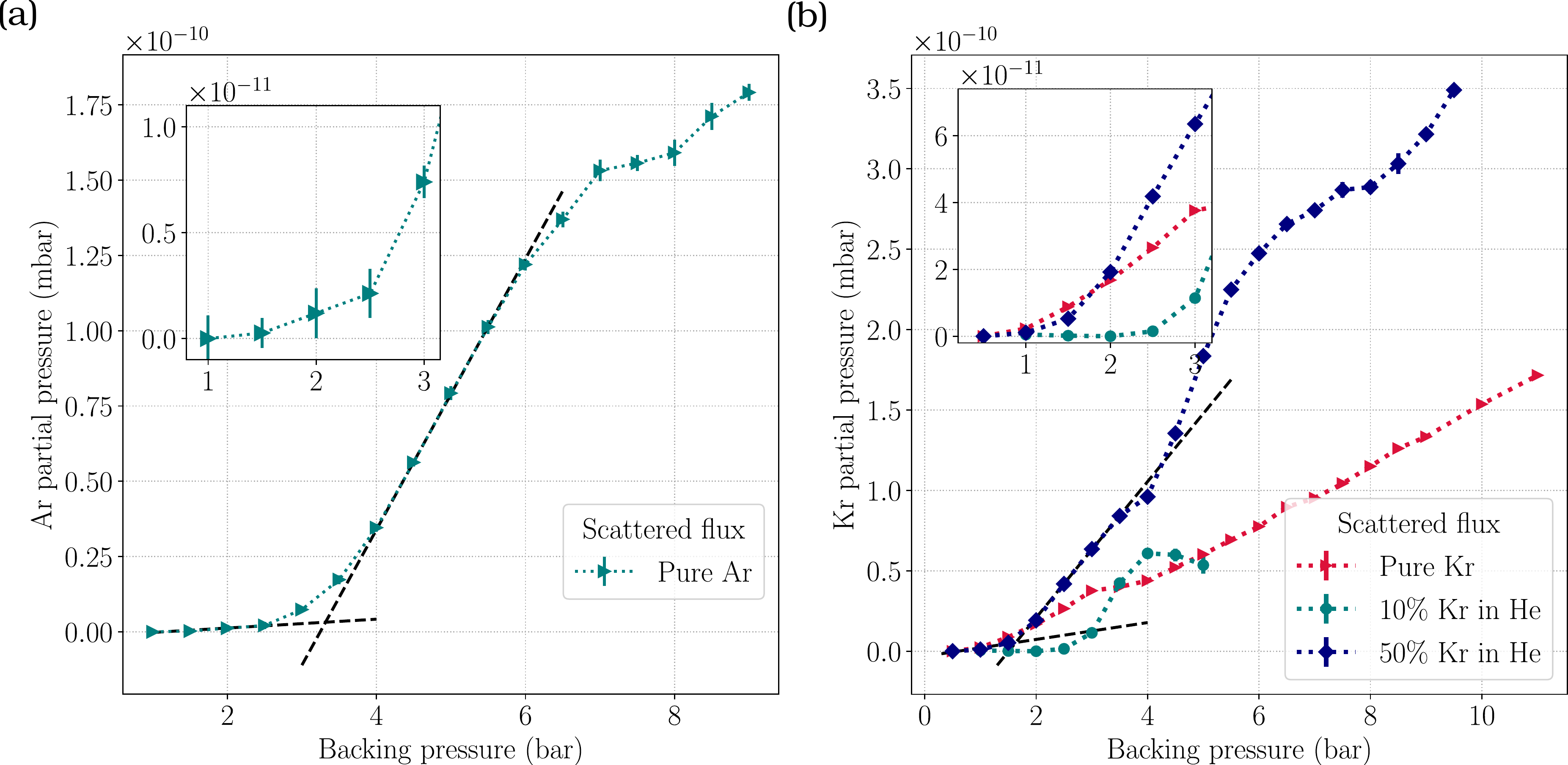}
\caption{Ar (a) and Kr (b) signal along center-line with increasing backing pressure. These measurements were made by monitoring the scattered signal from a rough surface. The solid lines show the linear fits in different pressure ranges.}
\label{fig:Ar_Kr_intensity_panel}
\end{figure}

Centerline intensity of a molecular beam is related to the backing pressure by the following relation:
\begin{equation}
I_0 \propto \dfrac{P_{0}d^2}{\sqrt{T_0}}
\end{equation}
Here, d is the orifice diameter, $P_0$ and $T_0$ are the stagnation pressure and temperature, respectively. 
According to the above relation, Kr signal is expected to increase linearly with increasing backing pressure.
However, we see a change in the slope for Ar at 3 bar and Kr beams in the range of 1 - 2.5 bar (Fig. \ref{fig:Ar_Kr_intensity_panel}).
We infer this change in slope to be caused by the onset of formation of large clusters, as suggested by the Hagena parameter.
Further evidence for large cluster formation was obtained by performing femtosecond laser ionization, leading to characteristic X-ray emmission being observed for Ar and Kr.

\newpage

\section*{SI-3: Experimental setup to verify cluster formation using X-ray generation}

\begin{figure}[H]
\centering
\includegraphics[width = 0.7\linewidth, draft=false]{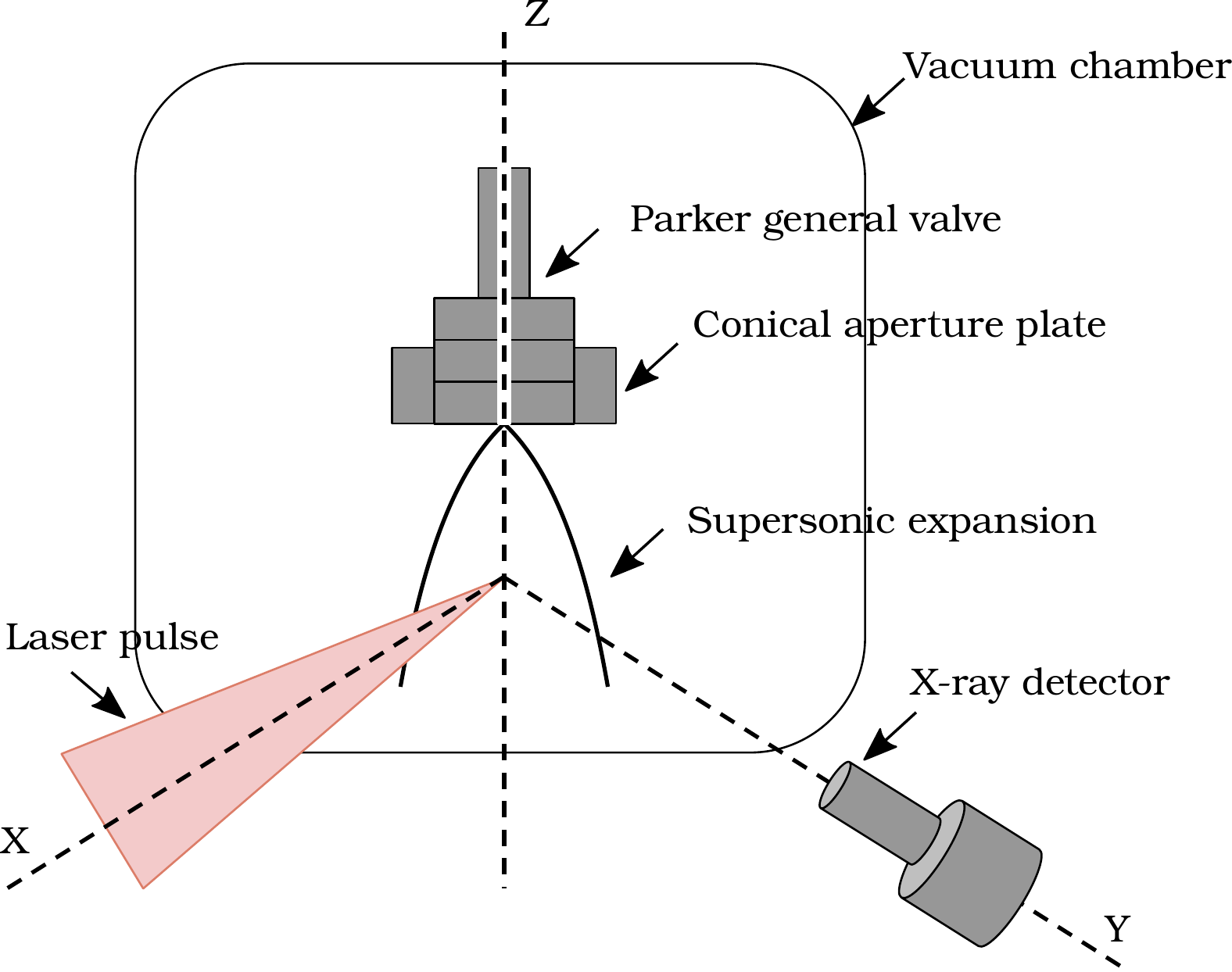}
\caption{Schematic diagram of the experimental setup used to study the  femtosecond laser ionization and consequent X-ray generation from  the beams of Ar and Kr clusters.}
\label{fig:xray_setup}
\end{figure}

Experimental setup used for X-ray generation and detection from Ar and Kr clusters is depicted in Figure \ref{fig:xray_setup}.
These measurements were performed with the same pulsed nozzle (with similar operating parameters) as used for NAM.
Focused pulses with a temporal width 25 fs (at 800 nm) having an energy of 2 mJ/pulse (intensity $\sim$ 10$^{16}$ W/cm$^2$) were made incident at the center of supersonic expansion, approximately 5 mm away exit plane.
This interaction generated X-rays which were detected using a Si-PIN detector (Amptek XR-100CR).
A significant amount of X-ray emission was observed under these conditions, clearly indicating that large clusters with sizes of the order of 10$^4$ atoms were formed for both Ar and Kr. 
\newpage

\section*{SI-4: X-ray spectra }

\begin{figure}[H]
\centering
\includegraphics[width = 1  \linewidth, draft=false]{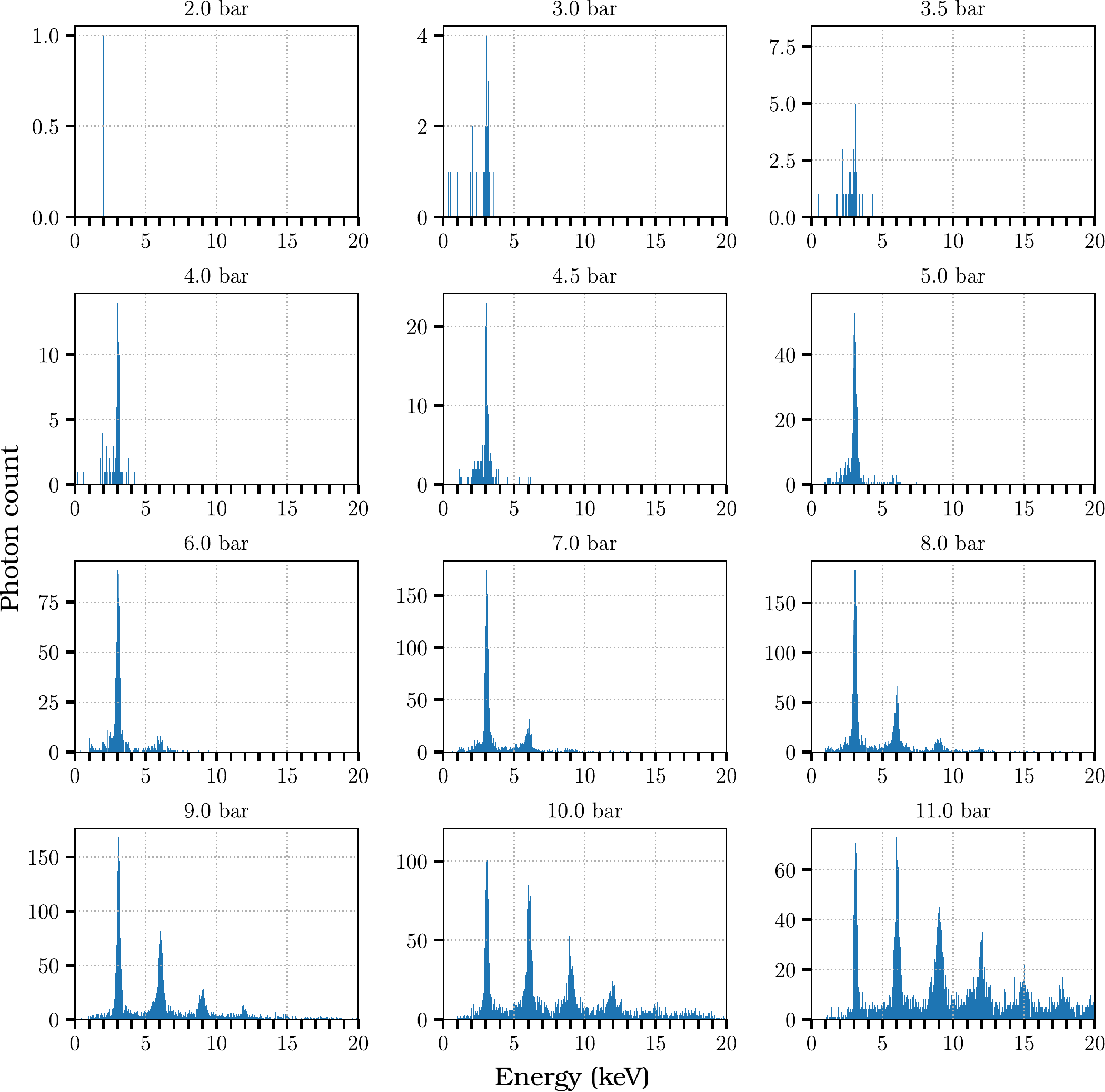}
\caption{Spectrum of the X-rays generated from Ar clusters interacting with intense fs laser pulses, measured at different backing pressures. Actual transition is at an energy of 3 keV. Peaks at 6, 9, 12, 15 and 18 keV energies merely indicate the accumulation of 2, 3, 4, 5 and 6 x-ray photons within the deadtime of detector, respectively.}
\label{fig:xray_Ar_panel}
\end{figure}

\begin{figure*}[t]
\centering
\includegraphics[width = 1\linewidth, draft=false]{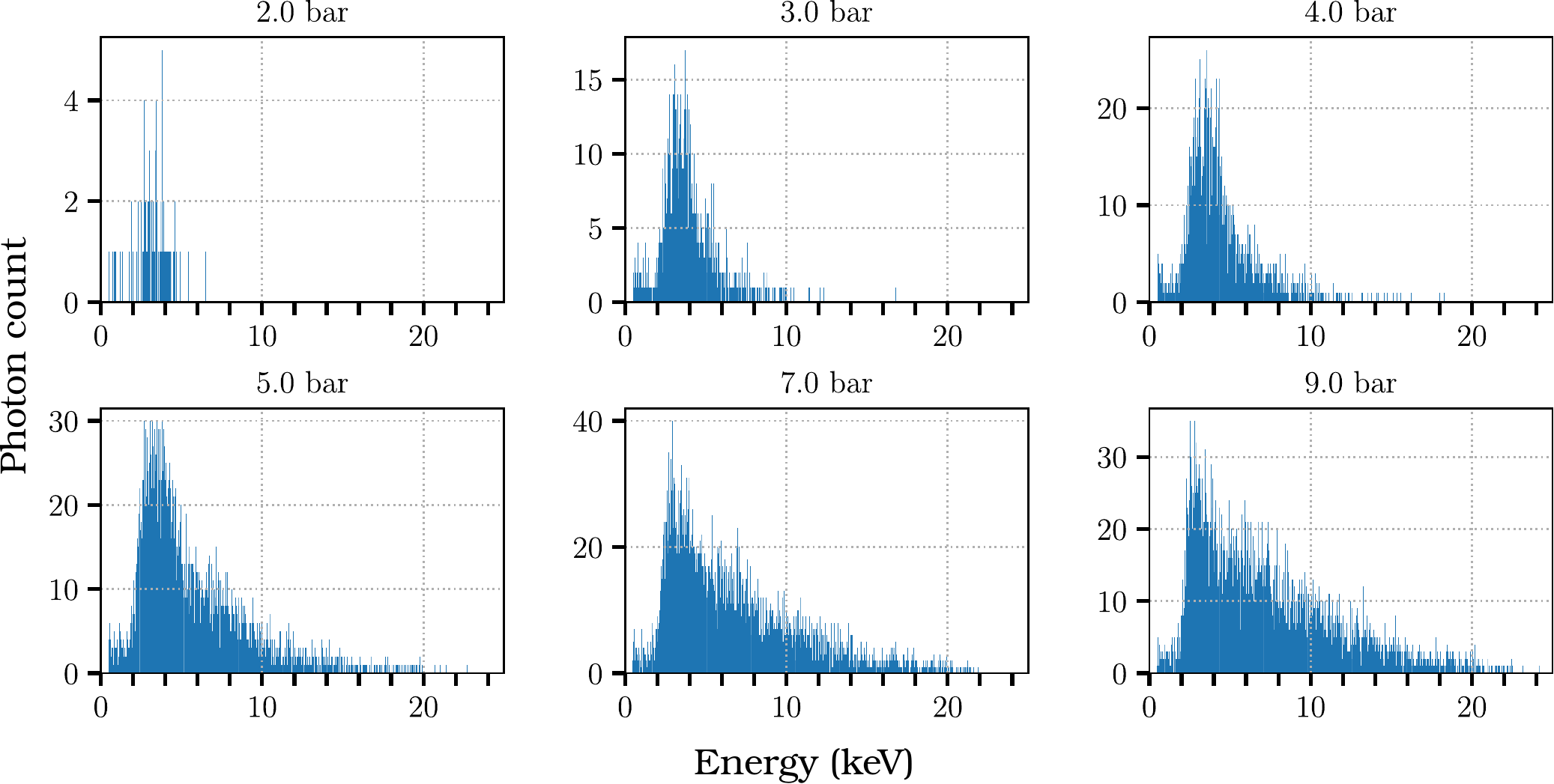}
\caption{Spectrum of the X-rays generated from Kr clusters interacting with intense fs laser pulses, measured at different backing pressures (pure Kr). Here we do not observe the separated peaks as in case of Ar. Rather a continuum (inverse bremsstrahlung) was observed due to the as expected for larger atoms.}
\label{fig:xray_Kr_panel}
\end{figure*}
